\documentclass[runningheads]{llncs}
\usepackage[T1]{fontenc}
\usepackage{graphicx}
\usepackage{enumitem}
\usepackage{hyperref} 
\usepackage{color}
\usepackage{listings}
\usepackage{xcolor}

\begin{document}
\title{Enhancing Type Safety in MPI with Rust: A Statically Verified Approach for RSMPI}
\titlerunning{Enhancing Type Safety in MPI with Rust}
%
\author{Nafees Iqbal\inst{}\orcidID{0009-0006-7712-2016} \and
Jed Brown\inst{}\orcidID{0000-0002-9945-0639}}
%
%
\institute{University of Colorado Boulder, USA\\
\email{\{nafees.iqbal,jed.brown\}@colorado.edu}}
\maketitle              
\begin{abstract}
The Message Passing Interface (MPI) is a fundamental tool for building high-performance computing (HPC) applications, enabling efficient communication across distributed systems. Despite its widespread adoption, MPI’s low-level interface and lack of built-in type safety make it prone to runtime errors, undefined behavior, and debugging challenges, especially in large-scale applications. Rust, a modern systems programming language, offers a compelling solution with its strong type system, which enforces memory and type safety at compile time without compromising performance. This paper introduces a type-safe communication framework for MPI, built on the RSMPI library, to address the limitations of traditional MPI programming. At its core is the \texttt{TypedCommunicator}, an abstraction that enforces static type safety in point-to-point communication operations. By leveraging Rust’s \\ \texttt{Equivalence} trait, our framework guarantees that only compatible types can participate in communication, catching mismatches either at compile time or through runtime validation. The framework supports both single-value and slice-based communication, providing an intuitive API for diverse data structures. Our implementation demonstrates that this approach eliminates common MPI errors, improves developer productivity, and maintains performance, adhering to Rust’s principle of zero-cost abstractions. This work lays the foundation for extending type safety to collective operations, advancing the robustness of parallel computing in Rust.

\end{abstract}
\lstdefinelanguage{Rust}{
    morekeywords=[1]{let, fn, mut, struct, impl, for, in, if, else, while, loop, match, pub, use, crate, enum, mod, self, super, const, static, trait, as, ref, break, continue, return, move, async, await, unsafe, new, TypedCommunicator},
    morekeywords=[2]{f32, i32}, 
    sensitive=true,
    morecomment=[l]{//},        
    morecomment=[s]{/*}{*/},   
    morestring=[b]{"},         
}

\lstset{
    language=Rust,
    basicstyle=\ttfamily\small,       
    keywordstyle=[1]\color{blue}\bfseries, 
    keywordstyle=[2]\color{teal},    
    commentstyle=\color{gray}\itshape, 
    stringstyle=\color{red},         
    numbers=left,                    
    numberstyle=\tiny\color{gray},   
    stepnumber=1,                    
    backgroundcolor=\color{lightgray!20}, 
    frame=single,                    
    showstringspaces=false,          
    breaklines=true,                 
    captionpos=b,                    
}

\section{Introduction}
Distributed parallel computing is essential for modern scientific and technological advancements, enabling applications that demand substantial computational resources. The Message Passing Interface (MPI)~\cite{mpi_standard} remains ubiquitous in distributed systems due to its flexibility, library-friendliness, and efficiency in high-performance computing (HPC)~\cite{lopez2014survey}~\cite{peisert2010fingerprinting}. However, MPI’s low-level interface lacks enforced type safety, requiring developers to manually ensure correct data alignment between sending and receiving processes. This absence of built-in safety mechanisms increases the risk of type mismatches, undefined behavior, and hard-to-debug runtime errors, particularly in large-scale applications~\cite{mpi_standard}.

Rust~\cite{rust_documentation} is a modern systems programming language that emphasizes memory safety and type safety, which are critical for developing reliable software. Rust's type system prevents common programming errors such as dangling pointers and data races, prevalent in languages like C and C++ ~\cite{jung2017rustbelt}~\cite{jung2019stacked}. Through ownership, borrowing, and lifetimes, Rust enforces static type safety at compile time, ensuring well-typed programs do not exhibit undefined behaviors that can lead to security vulnerabilities~\cite{astrauskas2019leveraging} and hard-to-debug behavior. These characteristics make Rust a compelling choice for addressing the safety concerns inherent in MPI programming, where the lack of robust safeguards can lead to hard-to-debug~\cite{tronge2023improving} issues.

RSMPI~\cite{rsmpi} is a library designed to bridge Rust and MPI, providing essential functionality for distributed and parallel computing in Rust. While RSMPI offers partial safety through Rust's memory-safe abstractions, it falls short in key areas such as type matching and validation during collective operations~\cite{tronge2023improving}. These mismatches can lead to wrong program behavior, or application crash issues that are particularly costly in high-performance computing (HPC) environments. By addressing these vulnerabilities, this project seeks to mitigate a significant source of unsoundness in MPI applications.

In this work, we focus on enhancing RSMPI by adding static type safety, especially for point-to-point communication. Static type safety ensures that data types used in communication operations match correctly at compile time, preventing errors before the program even runs. This eliminates a major source of erros caused by type mismatches during data exchange between processes. A key concept here is congruence, which means that the data types being communicated must align in both size and representation. 
Our approach introduces the \texttt{TypedCommunicator}, an abstraction that enforces compile-time type safety while preserving MPI’s performance. Unlike runtime validation, which catches errors only during execution, static type checking identifies type mismatches at compile time, significantly reducing debugging complexity and enhancing developer productivity. By ensuring that only compatible types are used in communication, our approach eliminates subtle errors and strengthens the robustness of MPI applications.

In summary, our key contributions are as follows:
\begin{itemize}[label=$\bullet$]
    \item Introduced the \texttt{TypedCommunicator} abstraction that enforces static type safety in MPI operations, ensuring that only congruent data types are exchanged between processes.
    \item Utilized Rust’s \texttt{Equivalence} trait to ensure that only compatible data types can be sent or received, catching potential type mismatches at compile time or as runtime panics.
    \item Unified communication handling with a single send function leveraging the \texttt{Buffer} trait. This design supports both single values and slices seamlessly, simplifying the API while maintaining type safety.
\end{itemize}

\section{Design and Implementation}
This section outlines the design and implementation of a statically type-safe communication framework in Rust, built upon the existing RSMPI library. Our goal is to enhance type safety in MPI operations by leveraging Rust’s strong type system while maintaining flexibility and performance. Unlike traditional MPI programming, which relies on manual type validation or runtime checks, our framework enforces type correctness at compile time whenever possible, supplemented by runtime checks where necessary. The core of this implementation is the \texttt{TypedCommunicator}, which encapsulates safe and efficient communication mechanisms for point-to-point MPI operations. Leveraging Rust's robust type system, the \texttt{TypedCommunicator} enforces type congruence across all participating processes, reducing the potential errors in parallel applications. Below, we discuss the key components and features of the framework, highlighting its role in ensuring correctness and usability in MPI programming.
\subsection{Ensuring Compile-Time Safety}
The \texttt{TypedCommunicator} uses a generic type parameter \texttt{T} to represent the data type in communication operations. By requiring T to implement Rust’s \texttt{Equivalence} trait, we ensure that only compatible data types are allowed during initialization. This guarantees compile-time validation, preventing type mismatches early in the development process.
\begin{lstlisting}[language=Rust, caption={Enforcing type consistency at compile-time.}, label={lst:compile_time_validation}]
let typed_comm: TypedCommunicator<f32> = TypedCommunicator::new(world_ref);
if rank == 0 {
    let data_to_send: f32 = 42.5;
    typed_comm.send(&data_to_send, 1, 0);
    // This works as the type matches the communicator's expected type.
} else if rank == 1 {
    let mut data_to_receive: i32 = 0;
    // This will fail to compile because the communicator expects f32, not i32.
    typed_comm.receive(&mut data_to_receive, 0, 0);
}
\end{lstlisting}
As shown in listing~\ref{lst:compile_time_validation}, the \texttt{TypedCommunicator} is explicitly initialized for \texttt{f32}. In this example, the first branch type-checks \texttt{Rank 0} sending an \texttt{f32} value, which matches the communicator’s expected type. However, the \texttt{Rank 1} is a type error because the receiving type \texttt{i32} is not congruent with the communicator's type \texttt{f32}. This compile-time enforcement prevents type mismatches early in development, enhancing the robustness and safety of MPI applications. 

\subsection{Ensuring Type Congruence Across Ranks}
In MPI, congruence extends beyond simple type equality by ensuring that the underlying structure and representation of data are compatible across communicating processes. This flexibility allows for compatibility even when data layouts differ. Consider the following listing~\ref{lst:congruence}, a sending process transmits a 3x2 matrix, and the receiving process directly reconstructs it into a 2x3 matrix. Despite the layout difference, they remain congruent as the total data size and structure are compatible.
\begin{lstlisting}[language=Rust, caption={Congruence in MPI: flexible data layouts for communication}, label={lst:congruence}]
if world.rank() == 0 {
    let typed_comm: TypedCommunicator<f32> = TypedCommunicator::new(world_ref);
    let x = [[1.0_f32; 2]; 3];
    typed_comm.send(&x, 1, 0);
} else if world.rank() == 1 {
    let typed_comm: TypedCommunicator<f32> = TypedCommunicator::new(world_ref);
    let mut y = [[0.0_f32; 3]; 2];
    typed_comm.receive(&mut y, 0, 0);
}
\end{lstlisting}
However, some type mismatches cannot be detected at compile-time. Consider the following listing~\ref{lst:type_mismatch}, where Rank 0 sends an \texttt{f32} value, while Rank 1 expects an \texttt{i32}. This issue arises because \texttt{TypedCommunicator} instances are initialized within separate conditional blocks with inconsistent types. Rust's type inference cannot catch this mismatch at compile-time, as the type contexts are isolated. To mitigate such issues, explicitly declaring the type outside conditional blocks ensures consistency and enables compile-time validation. When this is not feasible, run-time type checks play a crucial role in detecting and preventing mismatches, enhancing the robustness and reliability of MPI applications. Note that, run-time checking is a one-time cost for each \texttt{TypedCommunicator}. In real applications, communicators are typically used many times, and our approach has no run-time overhead relative during use. 
\begin{lstlisting}[language=Rust, caption={Example of type mismatch in \texttt{TypedCommunicator} usage across ranks}, label={lst:type_mismatch}]
if world.rank() == 0 {
    let typed_comm: TypedCommunicator<f32> = TypedCommunicator::new(&world);
    let data: f32 = 42.0;
    typed_comm.send(&data, 1, 0);
} else if world.rank() == 1 {
    let typed_comm: TypedCommunicator<i32> = TypedCommunicator::new(&world);
    let mut buffer: i32 = 0;
    typed_comm.receive(&mut buffer, 0, 0);
}
\end{lstlisting}
 
\subsection{Communication Methods}
The \texttt{TypedCommunicator} simplifies MPI communication by unifying single-value and slice-based data handling through a single, versatile \texttt{send} method. This approach leverages the \texttt{Buffer} trait, which abstracts the underlying data types and ensures compatibility with MPI’s datatype requirements. By integrating the \texttt{Buffer} trait, the send method seamlessly supports various data structures, such as single values, arrays, and slices, while maintaining strict type safety.

\section{Related Work}
MPI, while a powerful standard for high-performance computing, historically lacks type safety. This can lead to runtime errors, performance bottlenecks, and debugging challenges. The most closely related work is by Tronge et al.~\cite{tronge2023improving}, who developed a prototype for safe point-to-point messaging by modifying MPI to carry type information in every message, detecting runtime errors but requiring a special MPI implementation. In contrast, our approach catches type mismatches at compile time, ensures correctness before execution, requires no MPI changes, and can be selectively applied. Traditional tools like the Intel Trace Analyzer and Collector (ITAC)~\cite{intel_trace_analyzer} and Marmot Umpire Scalable Tool (MUST)~\cite{hilbrich2013mpi} focuses on runtime validation, detecting errors during execution rather than preventing them at compile time.

Recent research highlights the need for modern MPI bindings. Ruefenacht et al.~\cite{ruefenacht2021mpis} propose a new C++ interface leveraging C++11-20 features for improved performance and productivity. Ghosh et al.~\cite{ghosh2021towards} discuss ongoing efforts to provide standardized C++ bindings, emphasizing compact abstractions. Gregor et al.~\cite{gregor2008mpi} advocate for better support of high-level language bindings in MPI, noting the rise of Python and Java in HPC. Hespe et al.~\cite{uhl2024kamping} present KaMPIng, a flexible C++ binding for MPI that offers multiple abstraction levels.

In the realm of programming languages, Rust has gained significant recognition for its strong focus on memory safety~\cite{wang2018krust} and concurrency. Blesel et al.~\cite{blesel2021heimdallr} introduced heimdallr, a Rust-based message-passing library emphasizing compile-time correctness checks. 
Yu et al.~\cite{yu2019fearless} provide insights into concurrency safety in Rust, which is essential for MPI applications involving concurrent processes.

Our work builds upon these foundations by introducing a statically type-safe MPI interface in Rust through the \texttt{TypedCommunicator} abstraction. Unlike runtime validation tools, our approach leverages Rust's strong type system to enforce type compatibility at compile time, ensuring that all participating processes use compatible and congruent data types.
\section{Conclusion}
This work addresses a critical gap in MPI programming by introducing a statically type-safe communication framework built on the RSMPI library. Through the \texttt{TypedCommunicator} abstraction, we enforce type correctness in point-to-point MPI operations, reducing type mismatches and undefined behavior. By integrating compile-time validation with runtime checks, our approach ensures reliable communication while maintaining flexibility for both single-value and slice-based data transfers. The framework enhances MPI usability without compromising performance, making it suitable for high-performance computing applications. While this work focuses on point-to-point communication, it lays the foundation for extending type safety to collective operations and broader parallel computing models, contributing to more robust and scalable distributed systems.
\bibliographystyle{splncs04}
\bibliography{biblio}

\end{document}